\documentclass[review,12pt]{elsarticle}

\usepackage{amssymb}
\usepackage{amsthm}
\usepackage{amsmath}
\usepackage{mathrsfs}
\usepackage{graphicx}
\usepackage{epstopdf}
\usepackage{float}
\usepackage{caption}
\usepackage{subcaption}
\usepackage{bm}
\usepackage{amssymb} 
\usepackage{extarrows} 
\usepackage{bbm}
\usepackage{mathrsfs}
\usepackage{hyperref}
\usepackage{cleveref}
\usepackage{nameref}
\usepackage{soul}
\usepackage{booktabs}
\usepackage[margin=2.2cm]{geometry}
\usepackage{tabu} 
\usepackage{longtable}
\usepackage{xcolor} 
\usepackage{enumitem}

\journal{International Journal of Hydrogen Energy}

\makeatletter
\def\@author#1{\g@addto@macro\elsauthors{\normalsize%
    \def\baselinestretch{1}%
    \upshape\authorsep#1\unskip\textsuperscript{%
      \ifx\@fnmark\@empty\else\unskip\sep\@fnmark\let\sep=,\fi
      \ifx\@corref\@empty\else\unskip\sep\@corref\let\sep=,\fi
      }%
    \def\authorsep{\unskip,\space}%
    \global\let\@fnmark\@empty
    \global\let\@corref\@empty  
    \global\let\sep\@empty}%
    \@eadauthor={#1}
}
\makeatother

\begin{document}

\begin{frontmatter}



\title{Computational predictions of hydrogen-assisted fatigue crack growth}


\author{Chuanjie Cui\fnref{Ox,IC}}

\author{Paolo Bortot\fnref{Tenaris}}

\author{Matteo Ortolani\fnref{Tenaris}}

\author{Emilio Mart\'{\i}nez-Pa\~neda\corref{cor1}\fnref{Ox,IC}}
\ead{emilio.martinez-paneda@eng.ox.ac.uk}

\address[Ox]{Department of Engineering Science, University of Oxford, Oxford OX1 3PJ, UK}

\address[IC]{Department of Civil and Environmental Engineering, Imperial College London, London SW7 2AZ, UK}

\address[Tenaris]{Tenaris, Dalmine 24044, Italy}


\cortext[cor1]{Corresponding author.}

\begin{abstract}
A new model is presented to predict hydrogen-assisted fatigue. The model combines a phase field description of fracture and fatigue, stress-assisted hydrogen diffusion, and a toughness degradation formulation with cyclic and hydrogen contributions. Hydrogen-assisted fatigue crack growth predictions exhibit an excellent agreement with experiments over all the scenarios considered, spanning multiple load ratios, H$_2$ pressures and loading frequencies. These are obtained without any calibration with hydrogen-assisted fatigue data, taking as input only mechanical and hydrogen transport material properties, the material’s fatigue characteristics (from a single test in air), and the sensitivity of fracture toughness to hydrogen content. Furthermore, the model is used to determine: (i) what are suitable test loading frequencies to obtain conservative data, and (ii) the underestimation made when not pre-charging samples. The model can handle both laboratory specimens and large-scale engineering components, enabling the Virtual Testing paradigm in infrastructure exposed to hydrogen environments and cyclic loading.  \\


\end{abstract}


\end{frontmatter}

\section*{Graphical abstract}
\begin{figure}[H]
\centering
\includegraphics[scale=0.26]{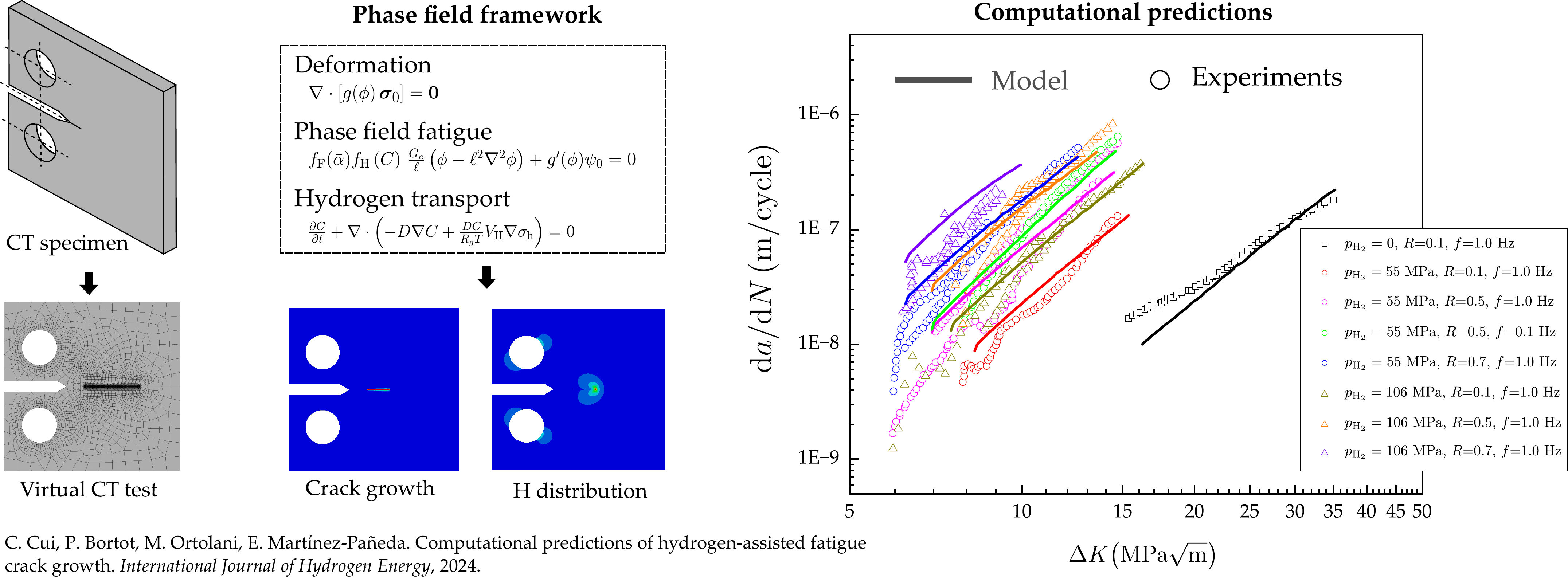}
\label{fig:abstract}
\end{figure}

\section{Introduction}
\label{Sec:Intro} 

\begin{figure}[H]
\centering
\includegraphics[scale=0.2]{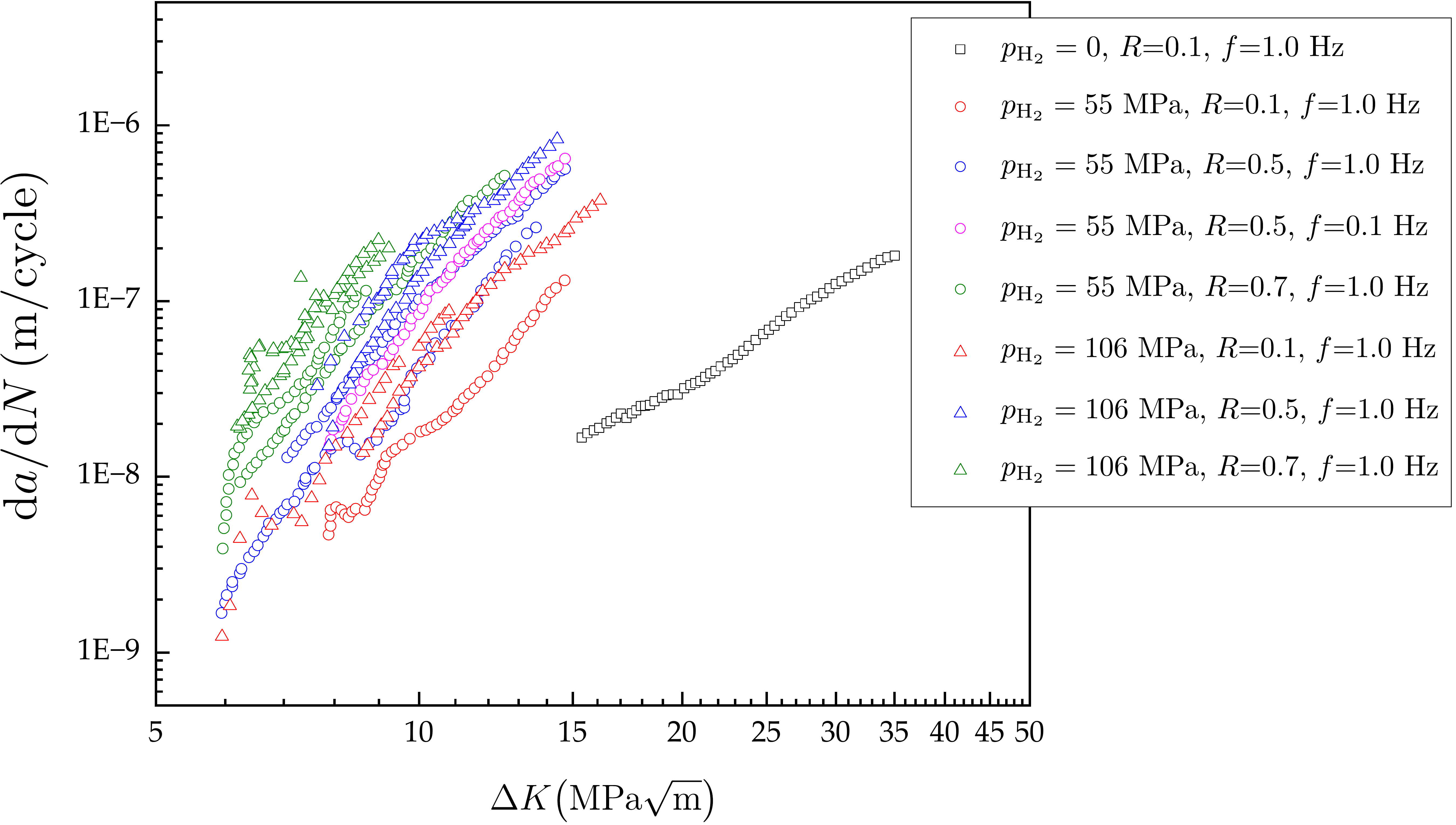}
\caption{Experimental fatigue crack growth rates d$a$/d$N$ versus stress intensity factor range $\Delta K$ curves for a ASME SA-723 Grade 1 pressure vessel steel \cite{San2017,bortot2023effect}. The experimental data have been generated for various hydrogen gas pressures ($p_\mathrm{H_2}$), load ratios ($R$), and loading frequencies ($f$). These tests, conducted at SANDIA National Laboratories, are representative of the fatigue crack growth behaviour of pressure vessel steels in hydrogen-containing environments. The curve in air ($p_\mathrm{H_2}=0$) was provided by TENARIS and is representative of the response in air of ASME SA-723 Grade 1 pressure vessel steel.} 
\label{fig:Expdata}
\end{figure}

Hydrogen is deemed to play a key role in decarbonisation strategies due to its abundance, eco-friendliness, diverse production sources, and multifaceted applications \cite{Petrollese2022,Makepeace2024,Rong2024}. However, contrary to this promise, hydrogen is famed for `embrittling' metallic materials, making its transport and storage notable technological challenges \cite{Gangloff2012hydrogen,Djukic2019,malheiros2022local}. The ingress of hydrogen into a metal can change - by orders of magnitude - its fracture toughness, ductility and fatigue crack growth resistance, through a phenomenon known as \emph{hydrogen embrittlement} \cite{Gangloff2003,RILEM2021}. The synergistic effects of hydrogen and fatigue damage are of particular interest, given that in many hydrogen energy storage and transport applications susceptible components are exposed to alternating mechanical loads \citep{Murakami2010a,Delbusto2017}. Experiments have revealed that hydrogen can modify the cyclic constitutive behaviour of pure metals and alloys \cite{Castelluccio2018,Hosseini2021}, as well as reduce the number of cycles required for the nucleation of cracks \cite{Esaklul1983,Esaklul1983b}. However, the most significant effect that hydrogen has under cyclic loading is a notable acceleration of fatigue crack growth rates \cite{Fernandez-Sousa2020,Shinko2021}. This is of pressing technological importance, as experiments on pipeline and pressure vessel steels have shown that fatigue crack growth rates can be orders of magnitude higher when samples have been exposed to hydrogen gas \cite{Nanninga2010, Briottet2015}. Figure \ref{fig:Expdata} shows representative results for pressure vessel steels \cite{San2017,bortot2023effect}, in terms of fatigue crack growth rate d$a$/d$N$ versus stress intensity factor range $\Delta K$ curves. These data have been obtained by testing an ASME SA-723 Grade 1, Class 1 steel\footnote{Equivalent, as per standard specification \cite{ASTMA372}, to the ASTM A372 Grade N, Class 100 steel.} \cite{ASMESA723} in various hydrogen gas environments (including the in-air condition, $p_\mathrm{H_2}=0$), under different load ratios ($R$) and loading frequencies ($f$). The data reveal the typical characteristics of hydrogen-assisted fatigue crack growth behaviour, being the most notable one the increase in fatigue crack growth rates with increasing hydrogen content, with experiments conducted in hydrogen gas environments showing values of d$a$/d$N$ that are more than ten times larger than those in air. Fatigue crack growth curves are also known to be sensitive to the loading frequency, as slow tests allow for more hydrogen to accumulate in the crack tip process zone \cite{Fernandez-Sousa2022}. Fatigue crack growth rates at low loading frequencies can be 5-10 times higher than those at high frequencies \cite{Suresh1982a,Tanaka2007,Kawamoto2008}, albeit smaller differences are observed in Fig. \ref{fig:Expdata}, where only two loading frequencies were considered (0.1 and 1 Hz). 

The data shown in Fig. \ref{fig:Expdata} are vital, as it enables fracture mechanics-based design of hydrogen storage and transport components. However, conducting cyclic loading experiments at high hydrogen gas pressures is not an easy task and is still only within the capabilities of a handful of laboratories. As such, there are significant benefits in developing a computational framework that can enable conducting reliable \emph{virtual} hydrogen-assisted fatigue crack growth experiments. For example, a mechanistic hydrogen-assisted fatigue crack growth model would enable to cheaply and efficiently produce new data, interpolating and extrapolating beyond the output of laboratory experiments, which is expensive and time-consuming to obtain and is tied to the specific conditions of the test (choices of material, $p_\mathrm{H_2}$, $f$, $R$). Moreover, computational predictions would allow addressing the `frequency conundrum' - while experiments are conducted at relatively high frequencies (0.1-1 Hz), due to practical constraints (duration, leak-prevention), engineering applications involve much lower loading frequencies ($f \ll 1$ Hz). In addition, a model capable of resolving the physical processes involved in hydrogen-assisted fatigue would enable taking into consideration the crack nucleation regime and delivering predictions for arbitrary loading histories.\\

In this work, we propose and validate a new model for predicting hydrogen-assisted fatigue. The model resolves the coupled deformation-diffusion-fracture/fatigue nature of the problem and employs a phase field approach to describe the evolution of cracks, building upon the growing body of literature showcasing the success of phase field models in the area of hydrogen embrittlement \cite{Martinez-Paneda2018,Duda2018,Anand2019,Isfandbod2021,Huang2020,JMPS2022,Dinachandra2022,Hageman2023,Si2024,Liu2024}. The results obtained show that the model can naturally predict hydrogen-assisted fatigue crack growth behaviour, taking as input solely the toughness versus hydrogen content dependency of the material and its fatigue response in air. Without any fitting or modification, the model is shown to \emph{blindly} estimate with remarkable accuracy fatigue crack growth behaviour across the entire range of load ratios, hydrogen pressures and loading frequencies considered. 
The validated model is then used to gain insight into the role of loading frequency and pre-charging times, mapping regimes of behaviour and assessing the importance of experimental constraints. Limitations and future enhancements are also discussed. The remainder of this paper starts by describing the coupled deformation-diffusion-fatigue phase field-based model presented, with additional details being given in the \nameref{app:Supplementary material}. Then, the numerical experiments are described and discussed. Finally, the manuscript ends with a brief summary and concluding remarks.


\section{A phase field-based model for hydrogen assisted fatigue}
\label{Sec:theory}

The formulation employed in this work is briefly presented here, with details of the numerical implementation given in \nameref{app:Supplementary material}. The model is effectively a combination of the state-of-the-art, generalised phase field fatigue model recently presented by Golahmar \textit{et al.} \cite{Golahmar2023} and the phase field formulations for hydrogen embrittlement by Mart\'{\i}nez-Pa\~neda and co-workers \cite{Martinez-Paneda2018,Golahmar2022,Kristensen2020}, whereby cracks evolve based on Griffith's energy balance, with the material toughness being degraded by both hydrogen and fatigue degradation functions. Throughout this work, we refer to an isotropic, elastic body occupying an arbitrary domain $\Omega \subset \mathbb{R}^{n}(n \in[1,2,3])$, with an external boundary $\partial \Omega \subset \mathbb{R}^{n-1}$, on which the outwards unit normal is denoted as $\mathbf{n}$. The basic framework of phase field fracture is introduced first, followed by its integration with hydrogen and fatigue damage. 

\subsection{Phase field regularisation of cracks}
\label{sec:PFR}

\begin{figure}[H]
\centering
\includegraphics[scale=0.6]{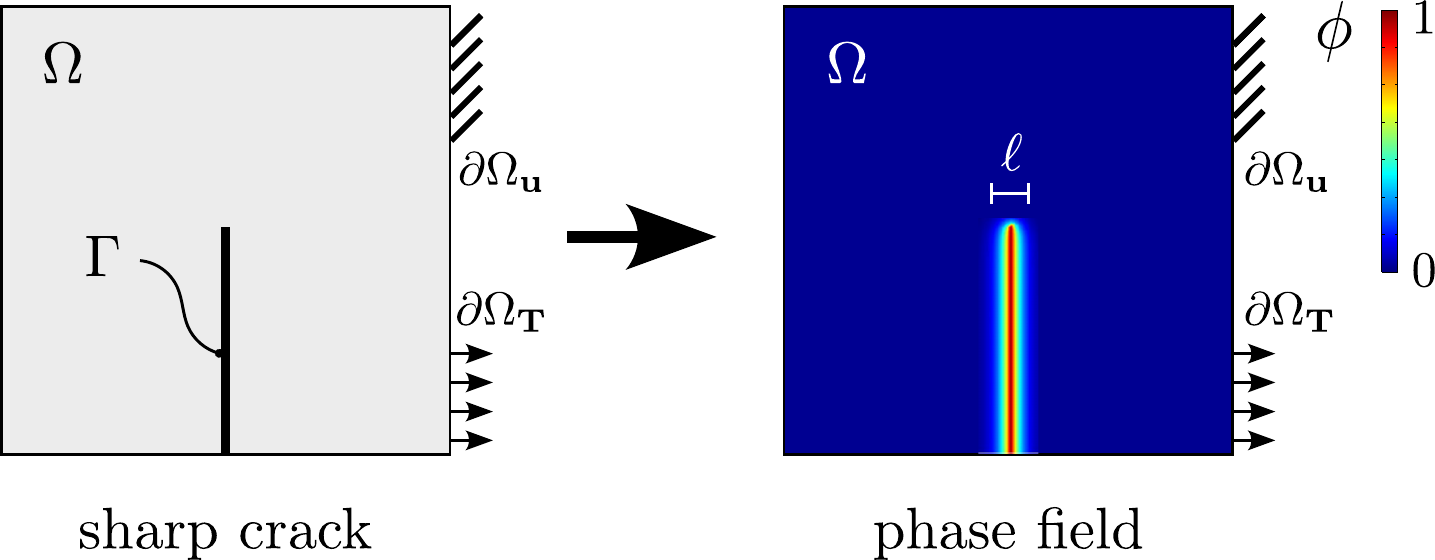}
\caption{Schematic description of phase field fracture. The discrete, sharp crack (left) is regularised by making use of an auxiliary phase field variable $\phi$, with the width of the regularised zone being governed by the magnitude of the phase field length scale $\ell$.} \label{fig: PFF}
\end{figure}

The phase field description of fracture is based on the thermodynamics of fracture, as first postulated by Griffith \cite{Griffith1920}. Griffith's theory describes fracture as a competition between the energy stored in the bulk and the energy required to create two new surfaces. In other words, crack propagation occurs when the energy stored in the material is sufficient to overcome the material toughness $G_c$. Denoting $\psi$ to be the density of elastic strain energy stored in the material and $\textbf{T}$ a traction acting over the surface of the body, the total energy functional $\Pi$ of the cracked solid can be written as,
\begin{equation}\label{eq:Griffith}
  \Pi=\int_{\Omega} \psi(\bm{\varepsilon}) \mathrm{~d}V + \int_{\Gamma} G_c \mathrm{~d}\Gamma -  \int_{\partial \Omega_T} \textbf{T} \cdot \textbf{u} \mathrm{~d}S,
\end{equation}
where $\mit\Gamma$ is the discrete crack surface and $\bm{\varepsilon}$ is the strain tensor, which is a function of the displacement field as $\bm{\varepsilon}=\left(\nabla \textbf{u} + \nabla \textbf{u}^T\right)/2$. Minimisation of Eq. (\ref{eq:Griffith}) would enable capturing the evolution of cracks as an exchange between elastic and fracture energies. However, this problem is computationally intractable since the crack surface $\mit \Gamma$ is usually unknown. Phase field modelling enables overcoming this problem by introducing an auxiliary variable, the phase field order parameter $\phi$, to track the crack interface \cite{Francfort1998,Bourdin2008}. As sketched in Fig. \ref{fig: PFF}, the order parameter $\phi$ varies smoothly from 0, at intact material points, to 1, in fully cracked regions, smearing the sharp crack into a diffuse region. Subsequently, the spatial and temporal evolution of the crack, including its nucleation, propagation, branching, and merging behaviours, becomes a natural by-product of Griffith's energy balance. Adopting the so-called conventional or \texttt{AT2} phase field model, the total potential energy (\ref{eq:Griffith}) can be approximated by
\begin{equation}\label{eq:PFF}
   \Pi = \Pi \left( \textbf{u}, \phi\right) \approx \int_{\Omega} \psi(\bm{\varepsilon}) \mathrm{~d}V + \int_{\Omega}  \frac{G_c}{2 \ell} \left( \phi^2 + \ell^2 \lvert \nabla \phi \rvert^2 \right) \mathrm{~d} V -  \int_{\partial \Omega_T} \textbf{T} \cdot \textbf{u} \mathrm{~d} S,
\end{equation}
with $\ell$ being the length scale parameter that governs the size of the smooth fracture process zone. Through $\Gamma$-convergence, it can be rigorously proven that the regularised function Eq. (\ref{eq:PFF}) converges to the Griffith's theory Eq. (\ref{eq:Griffith}) when $\ell \rightarrow 0^+$. This smeared description of fracture has enjoyed great success, with phase field formulations being developed for a number of fracture problems \cite{Liu2023,Feng2021,Feng2023,Korec2023}, from small-scale cracking in battery materials \cite{Ai2022,Parks2023} to crevasse growth in large ice sheets and glaciers \cite{Clayton2022,Sun2021}.

The balance equations can be readily derived by considering appropriate constitutive choices. First, following the usual assumption of a quadratic degradation function, the strain energy density in Eq. (\ref{eq:PFF}) can be given as,
\begin{equation}\label{eq:phiE}
\psi(\bm{\varepsilon})=\left( 1-\phi\right)^2 \psi_0(\bm{\varepsilon}),
\end{equation}
where $\psi_0(\bm{\varepsilon})$ is the undamaged strain energy density, which for elastic solids is given by
\begin{equation}\label{eq:phiE0}
   \psi_0(\bm{\varepsilon})= \frac{1}{2} \bm{\varepsilon} : \textit{\textbf{C}}_0 : \bm{\varepsilon},
\end{equation}
with $\textit{\textbf{C}}_0$ being the (undamaged) linear elastic stiffness tensor, as defined by appropriate choices of Young's modulus $E$ and Poisson's ratio $\nu$. Since the focus is on fatigue crack growth under small-scale yielding conditions, plasticity effects can be neglected. Nevertheless, extending the present framework to incorporate plasticity is relatively straightforward (see Ref. \cite{Mandal2024}). Consistent with Eq. (\ref{eq:phiE0}), the stress tensor is given by $\bm{\sigma}= g(\phi)\,\bm{\sigma}_0 =\left( 1-\phi\right)^2 \partial \psi_0(\bm{\varepsilon})/\partial \bm{\varepsilon}$, with $\bm{\sigma}_0$ being the undamaged stress tensor. Considering these constitutive choices, one can derive the governing equations by taking the stationary of Eq. (\ref{eq:PFF}) with respect to the primary fields, $\mathbf{u}$ and $\phi$, and applying the divergence theorem; such that,
\begin{equation}\label{eq:strongform}
\begin{aligned}
    \nabla \cdot\left[g(\phi)\, \bm{\sigma}_{0}\right] & =\mathbf{0} & \text { in } & \Omega \\
    \frac{G_{c}}{\ell} \left(\phi - \ell^2 \nabla^2 \phi\right)+ g^{\prime}(\phi)\,\psi_{0}(\bm{\varepsilon}) & =0 & \text { in } & \Omega
\end{aligned}
\end{equation}
with the following associated boundary conditions,
\begin{equation}\label{eq:strongform2}
\begin{aligned}
   g(\phi)\, \bm{\sigma}_0 \cdot \textbf{n} & = \textbf{T} & \text { on } & \Omega_T \\
   \nabla \phi \cdot \textbf{n} &=0 & \text { on } & \Omega.
\end{aligned}   
\end{equation}

Finally, it should be noted that the choice of $\ell$ defines the material strength $\sigma_c$ \cite{Tanne2018,Kristensen2021}. This can be readily observed by considering the analytical solution for Eq. (\ref{eq:strongform}b) in the case of a homogeneous 1-D scenario ($\nabla \phi=0$), where $\psi_0=E \varepsilon^2/2$ and the stress and strain distributions reach a maximum at
\begin{equation}\label{eq:sigmaAT2}
   \sigma_{c}=\frac{9}{16} \sqrt{\frac{E G_{c}}{3 \ell}} \, \, \, \,\, \,\, \,\, \,\, \,\varepsilon_c = \sqrt{\frac{G_c}{3 \ell E}}
\end{equation} 

\subsection{Hydrogen effects}
\label{sec:HE}

Let us now consider the role of hydrogen. To this end, one must consider the diffusion of hydrogen within the metal, which governs the loading frequency effect. Hydrogen transport is governed by mass conservation, such that the rate of change in time ($t$) of the hydrogen concentration ($C$) must be equal to the concentration flux $\mathbf{J}$ through the boundary of the body,
\begin{equation}\label{eq:Hydrogen_trans1}
   \int_{\Omega} \frac{\partial C}{\partial t} \mathrm{~d} V+\int_{\partial \Omega} \mathbf{J} \cdot \mathbf{n} \mathrm{~d} S=0.
\end{equation}
Since mass conservation must hold for any volume, Eq. (\ref{eq:Hydrogen_trans1}) can be re-written in a point-wise form as
\begin{equation}\label{eq:Hydrogen_trans2}
   \frac{\partial C}{\partial t}+\nabla \cdot \mathbf{J}=0.
\end{equation}

The hydrogen flux $\mathbf{J}$ is given by
\begin{equation}\label{eq:Jmu}
   \mathbf{J}=-D \nabla C + \frac{D C}{R_g T} \bar{V}_{\mathrm{H}} \nabla \sigma_\mathrm{h},
\end{equation}
where $D$ is the (apparent) hydrogen diffusion coefficient, $R_g$ is the gas constant, $T$ is the absolute temperature, $\bar{V}_\mathrm{H}$ is the partial molar volume of hydrogen in the solid solution, and $\sigma_\mathrm{h}$ is the hydrostatic stress. The last term in Eq. (\ref{eq:Jmu}) account for the effects of mechanical straining on the increase in hydrogen solubility, a result of lattice distortion.

Inserting Eq. (\ref{eq:Jmu}) into Eq. (\ref{eq:Hydrogen_trans2}) renders the governing equation for hydrogen transport,
\begin{equation}\label{eq:Htransport}
   \frac{\partial C}{\partial t}+ \nabla \cdot \left(-D \nabla C + \frac{D C}{R_g T} \bar{V}_{\mathrm{H}} \nabla \sigma_\mathrm{h}\right)=0.
\end{equation}

It remains to define the coupling between the hydrogen diffusion and fracture problems. Since hydrogen reduces the toughness of metals, this is commonly done through the definition of a hydrogen damage function $f_\mathrm{H}\left(C\right)$, such that the fracture energy becomes $f_\mathrm{H}\left(C\right) G_c$. Two approaches are generally taken to define $f_\mathrm{H}\left(C\right)$. A mechanistic one, whereby a degradation law is defined based on a multi-scale interpretation of the underlying mechanisms (e.g., quantitatively degrading $G_c$ via first-principles calculations \cite{Martinez-Paneda2018}). Or a phenomenological one, whereby the toughness sensitivity to hydrogen content is characterised experimentally \textit{via} fracture toughness tests at various hydrogen contents. Here, the latter is used since, as described below, significant fracture toughness versus hydrogen content data exist for pressure vessel steels. These data are best fitted through an exponential degradation law of the form \cite{Mandal2024},
\begin{equation}\label{eq:fc2}
   f_\mathrm{H}\left(C\right)=\xi+\left(1-\xi\right) \exp{\left(-\eta C^b \right)},
\end{equation}
where $\xi$, $\eta$, and $b$, the parameters that govern the extent of the fracture toughness degradation due to hydrogen, are to be calibrated with experiments. 

\subsection{Fatigue effects}
\label{sec:Fatigue}

The last ingredient of the model is the consideration of fatigue damage, and its couplings. The extension of phase field fracture to fatigue adopted here follows the framework established by Golahmar \textit{et al.} \cite{Golahmar2023}, which is itself based on the work by Carrara \textit{et al.} \cite{Carrara2020} and extends it to capture arbitrary S-N curve slopes, the fatigue endurance limit, and the mean stress effect. Accordingly, the material toughness is degraded by a fatigue degradation function $f_\mathrm{F}(\bar{\alpha})$, which evolves based on a cumulative history variable $\bar{\alpha}$, which is expressed as a function of a variable $\alpha$ to capture the physics of fatigue damage. Consistent with the fracture energy balance discussed above, fatigue damage is assumed to be driven by the energy stored in the solid and accordingly, the parameter $\alpha$ is defined as $\alpha=\psi$. The fatigue degradation function is given by \cite{Carrara2020}
\begin{equation}\label{eq:falpha}
    f_\mathrm{F}(\bar{\alpha})=\left(1-\frac{\bar{\alpha}}{\bar{\alpha}+\bar{\alpha}_{0}}\right)^{2},
\end{equation}
where $\bar{\alpha}_0$ is a material parameter determining the rate at which hydrogen affects the degradation of fracture toughness, to be calibrated with experiments. The fatigue history variable $\bar{\alpha}$ at a time $t$ is given by
\begin{equation}\label{eq:baralpha}
    \bar{\alpha}_{t}=\bar{\alpha}_{t-\Delta t}+\int_{t-\Delta t}^{t} \dot{\bar{\alpha}} \mathrm{~d} \tau=\bar{\alpha}_{t-\Delta t}+\Delta \bar{\alpha}.
\end{equation}

And the definition of $\Delta \bar{\alpha}$ is a key aspect of the model. Here, following Golahmar \textit{et al.} \cite{Golahmar2023}, we use a generalised definition of $\Delta \bar{\alpha}$ that accounts for (i) the exponent term of the Paris law, (ii) the effect of the stress ratio, and (iii) the fatigue threshold. Specifically, $\Delta \bar{\alpha}$ is defined as
\begin{equation}\label{eq:delta_baralpha}
    \Delta \bar{\alpha}=\left(\frac{\alpha_{\max}}{\alpha_{n}}\right)^{n}\left(\frac{1-R}{2}\right)^{2 \kappa n} H\left(\max _{\tau \in[0, t]} \alpha_{\max }\left(\frac{1-R}{2}\right)^{2 \kappa}-\alpha_{e}\right),
\end{equation}
where $n$ is a material parameter to match the slope of the S-N curve, $\alpha_{\max}$ is the maximum value of $\alpha$ within a fatigue cycle, $\alpha_{n}=\sigma_{c} \varepsilon_{c}/2$ is a normalisation parameter to achieve dimensional consistency, $R$ is the stress (load) ratio, $\kappa$ is a material constant measuring the sensitivity of fatigue damage to mean stress effect, $H()$ is the Heaviside function and $\alpha_{e}$ is the fatigue threshold corresponding to the material endurance stress. As per Ref. \cite{Golahmar2023}, the exponent term $n$ can be related to the exponent $m$ in the Paris law equation, $\mathrm{d}a/\mathrm{d}N=C (\Delta K)^m$, as $n=0.49 m -0.61$. \\

Considering both the fatigue and hydrogen degradation functions, the phase field evolution equation (\ref{eq:strongform}b) could be reformulated as\footnote{While this is the simplest form of the phase field evolution equation, it should be noted that if the toughness degradation is introduced in the weak form, gradient terms for the fatigue and hydrogen degradation functions naturally appear in the balance equation. Here, we follow Golahmar \textit{et al.} \cite{Golahmar2022} and consider only the former, neglecting the role of $\nabla f_\mathrm{H} (C)$.},
\begin{equation}\label{eq:strongformP}
    f_\mathrm{F}(\bar{\alpha}) f_\mathrm{H}\left(C\right)\,\frac{G_{c}}{\ell} \left(\phi - \ell^2 \nabla^2 \phi\right)+g^{\prime}(\phi) \psi_0 =0,
\end{equation}

Hence, both fatigue damage and hydrogen are assumed to degrade the material toughness synergistically. Importantly, we assume that there is no other coupling between the accumulation of hydrogen and the fatigue process. Fatigue crack growth rates will exhibit sensitivity to hydrogen content only through the hydrogen sensitivity of the material toughness. This reduces the number of parameters at play and is consistent with experimental observations - hydrogen significantly degrades the fracture toughness and accelerates fatigue crack growth rates but has a limited impact on the accumulation of fatigue damage, as evidenced by the relatively small effect on S-N curves. However, a more general version of our model is presented in \ref{app:AppendixX}, where the influence of hydrogen in accelerating the ability of a material to experience cyclic damage is accounted for. This leads to a slightly better agreement with experiments at the cost of having one additional parameter in the model. 

\section{Numerical fatigue crack growth experiments}
\label{Sec:Results}

The coupled phase field model presented is now used to conduct \emph{virtual} hydrogen-assisted fatigue crack growth experiments. In the following, we start by describing the testing configuration and choices of model parameters. Subsequently, numerical experiments are conducted to demonstrate the ability of the model to predict, without any fitting, the role of hydrogen gas pressure, load ratio and loading frequency. The validated model is then used to map the role of the loading frequency and assess the importance of the hydrogen exposure time prior to mechanical loading. 

\subsection{Preliminaries: testing configuration and material properties}

\begin{figure}[H]
\centering
\includegraphics[scale=0.8]{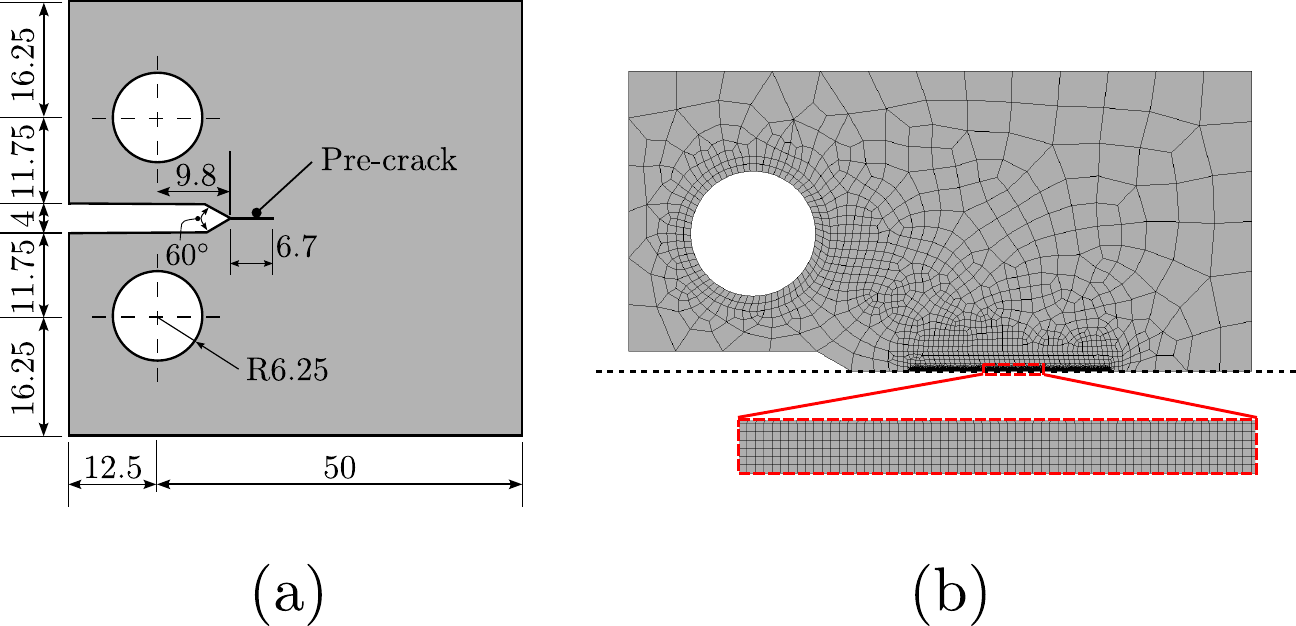}
\caption{Testing configuration; Compact Tension (CT) samples are employed with the following: (a) geometry, with dimensions in mm, and (b) finite element mesh, with the zoomed region showing the mesh along the expected crack growth trajectory.} \label{fig: CTspecimen}
\end{figure}

Mimicking the laboratory experiments, simulations are conducted using Compact Tension (CT) specimens with the dimensions given in Fig. \ref{fig: CTspecimen}. Taking advantage of symmetry, only the upper half of the sample is modelled. The finite element mesh, shown in Fig. \ref{fig: CTspecimen}b, employs approximately 5,000 plane strain quadratic quadrilateral elements, with the mesh being particularly fine in the expected crack growth region. To ensure mesh-independent results, the characteristic element size is taken to be at least 6 times smaller than the phase field length scale $\ell$. A cyclic load $P$, of range $\Delta P$, is applied as a nodal force on the pin. Following the ASTM E647 standard \cite{ASTME647}, the stress intensity factor range $\Delta K$ is then obtained from the applied load using the following expression,
\begin{equation}\label{eq:DeltaK}
  \Delta K=\frac{\Delta P\left(2+\frac{a}{W}\right)\left(0.886+4.46 \frac{a}{W}-13.32\left(\frac{a}{W}\right)^{2}+14.72\left(\frac{a}{W}\right)^{3}-5.6\left(\frac{a}{W}\right)^{4}\right)}{B \sqrt{W}\left(1-\frac{a}{W}\right)^{1.5}},
\end{equation}
where $a$ is the current crack length, $W$ is the effective width, and $B$ is the sample thickness.

Consistent with the experimental data, we simulate the behaviour of a pressure vessel steel, referred to as ASME SA-723 Grade 1, Class 1. This material has a Young's modulus of $E=210$ GPa and a Poisson's ratio of $\nu=0.3$. In the absence of hydrogen, the fracture behaviour of this material is described by a toughness of $G_c=100$ $\mathrm{kJ/m^2}$ \cite{SanMarchi2012}, and a critical strength that is four times the yield stress \cite{Mandal2024}, which is $\sigma_y=715$ MPa \cite{San2017}. Considering the relationship outlined in Eq. (\ref{eq:sigmaAT2})a, this results in a phase field length scale of $\ell=0.27$ mm. In terms of its fatigue behaviour, the model requires defining the parameters $n$, $\kappa$ and $\bar{\alpha}_0$. The sensitivity to the stress ratio is characterised by a choice of $\kappa=0.78$, based on the experiments conducted by Dowling \textit{et al.} \cite{Dowling2009} on a similar material. The parameter $n$, governing the slope of the Paris and S-N curves, is chosen to be $n=1.25$, to fit the slope observed in air. Finally, $\bar{\alpha}_0$ is taken to be equal to $\bar{\alpha}_0=8$ N/mm$^2$, so as to best capture the experimental fatigue crack growth curve in the absence of hydrogen. In this regard, we emphasise that the fatigue behaviour of the material in air ($p_\mathrm{H_2}=0$) is an input to the model (i.e., it requires parameter \emph{calibration}), unlike the fatigue behaviour in a hydrogen-containing environment, which is obtained without any changes to the model parameters and is, therefore, a \emph{prediction}. For completeness, we also define the fatigue threshold parameter, which is taken to be $\alpha_e=0.05$ (corresponding to an endurance stress of 150 MPa), although this parameter has no effect on the type of crack growth experiments conducted here \cite{Golahmar2023}.\\

Hydrogen is assumed to influence material behaviour only through the degradation of the material toughness $G_c$. To characterise the sensitivity of $G_c$ to hydrogen content, experimental fracture toughness versus hydrogen gas pressure data are fitted with a so-called hydrogen degradation function $f_\mathrm{H} (C)  = G_c / G_c (C=0) = J_{Ic}  / J_{Ic} (C=0)$, where $G_c (C=0)$ ($J_{Ic} (C=0)$) denotes the toughness in an inert environment. This function is an intrinsic material characteristic. To obtain $f_\mathrm{H} (C)$ from toughness versus $p_\mathrm{H_2}$ data, we first need to determine the hydrogen concentration corresponding to each of the hydrogen pressures considered in the experiments. To this end, Sievert's law is used \cite{Sieverts1929},
\begin{equation}\label{eq:Sievert}
    C=S\sqrt{{p_\mathrm{H_2}}},
\end{equation}
where the solubility of steel is defined as $S=0.077$ wppm $\mathrm{MPa}^{-0.5}$ \citep{Martin2020}. Under the sensible assumptions of small scale yielding and plane strain conditions, the toughness $G_c$ can be readily related to the critical $J$-integral or fracture toughness $K_{Ic}$ as,
\begin{equation}
    G_c = J_{Ic} = \frac{\left(1-\nu^2 \right) K_{Ic}^2}{E}
\end{equation}

\begin{figure}[H]
\centering
\includegraphics[scale=0.2]{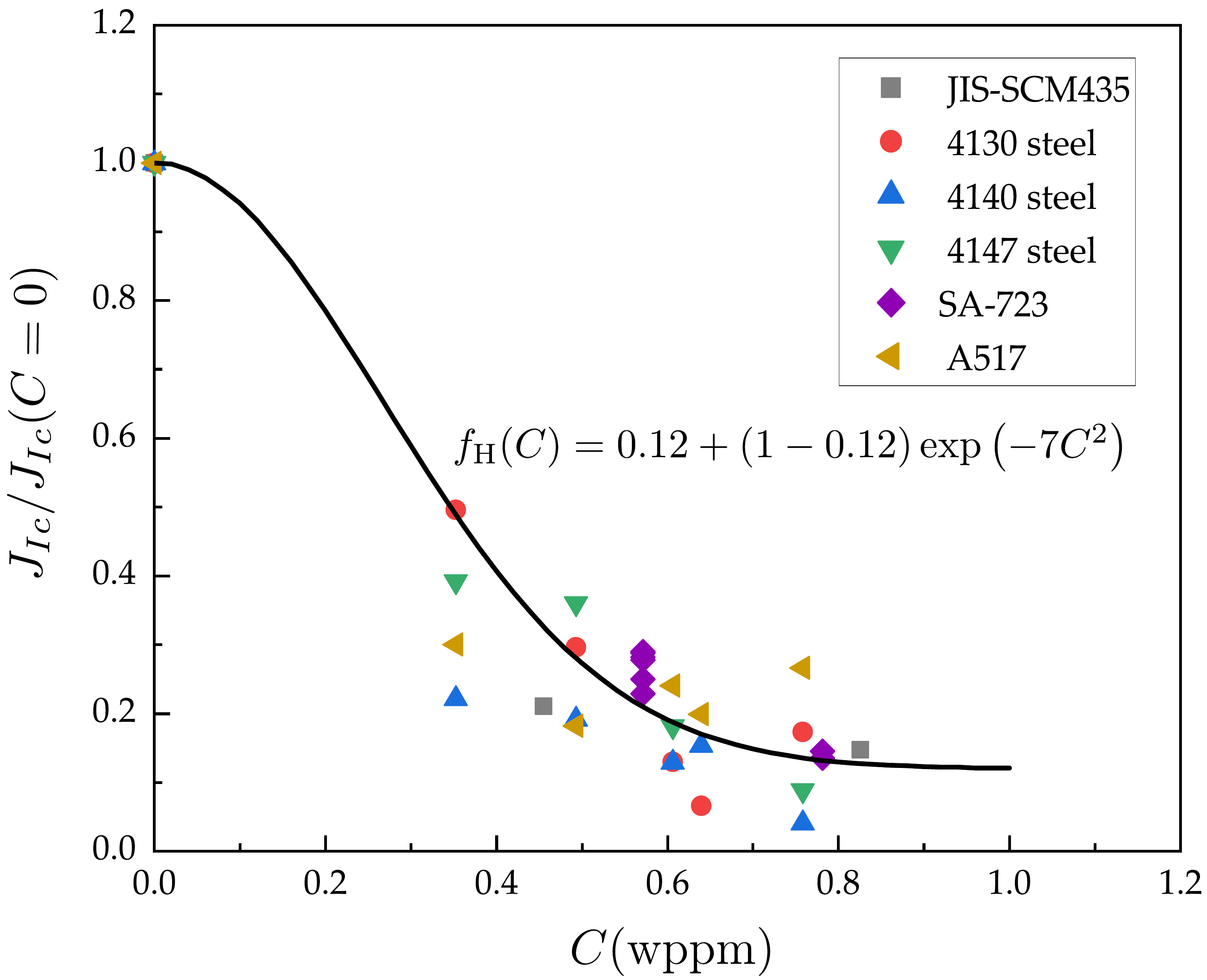}
\caption{Degradation of the fracture toughness as a function of the hydrogen concentration. Experimental data for similar pressure vessel steels (taken from Refs.   \cite{SanMarchi2012,Matsumoto2017,bortot2023effect,San2017}). The data are used to determine the material hydrogen degradation function $f_\mathrm{H} (C)  = G_c / G_c (C=0) = J_{Ic}  / J_{Ic} (C=0)$.} \label{fig:KthvsH}
\end{figure}

Accordingly, data for various hydrogen pressures can be condensed in a $J_{Ic}  / J_{Ic} (0)$ (or $f_\mathrm{H}$) versus $C$ plot, as shown in Fig. \ref{fig:KthvsH}, which uses (static) fracture toughness experimental data for similar pressure vessel steels. The data can be suitably fitted with a degradation function of the type given in Eq. (\ref{eq:fc2}), with $\xi=0.12$, $b=2$ and $\eta=7$.

Finally, the transport of hydrogen requires the definition of the partial molar volume, $\bar{V}_\mathrm{H}=2000$ mm$^3$/mol for steels, and the effective hydrogen diffusion coefficient, which is taken to be $D=2\times10^{-4}$ $\mathrm{mm^2/s}$, based on the range of experimental data reported \cite{Venezuela2018,Zafra2023}. The boundary conditions for the hydrogen transport problem mimic the conditions of the laboratory experiments. That is, starting from a hydrogen-free sample ($C(t=0)=0 \, \forall \, \mathbf{x}$), the outer surfaces of the specimen are exposed to the hydrogen concentration associated with the H$_2$ pressure used in the test, as per Sievert's law (\ref{eq:Sievert}). This environmental hydrogen content is referred to as $C_{\text{env}}$. As in the experiments, no mechanical load is applied in the first 24 h of hydrogen gas exposure (the `soaking' time associated with the stabilisation of the testing setup). Once the crack grows in a hydrogen-containing environment, we capture how the newly created crack surfaces are promptly exposed to the hydrogen environment by using a so-called penalty boundary condition \cite{Renard2020,CS2020}, which enforces $C=C_{\text{env}}$. \\

\begin{figure}[H]
\centering
\includegraphics[scale=0.5]{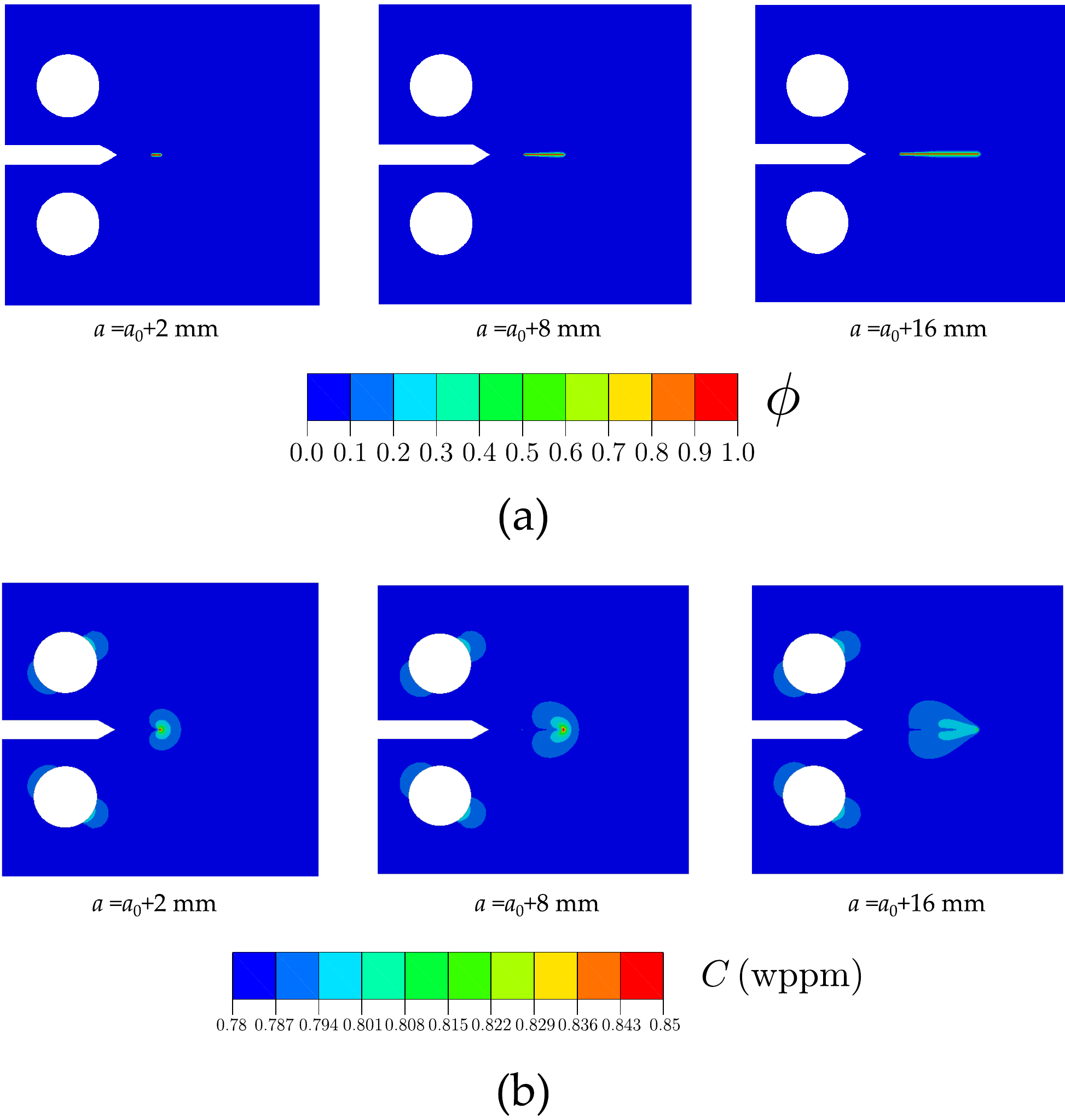}
\caption{Numerical fatigue crack growth experiments: representative results. Contours of (a) phase field variable $\phi$ and (b) hydrogen concentration $C$ with three different stages of crack growth: $a=a_0+2$ mm, $a=a_0+8$ mm, and $a=a_0+16$ mm, where $a_0$ is the initial crack size, as shown in Fig. \ref{fig: CTspecimen}a. These representative contours have been computed for a H$_2$ pre-charged sample exposed to a loading frequency of $f=1$ Hz, a hydrogen pressure of $p_\mathrm{H_2}=106$ MPa and a loading ratio of $R=0.1$.} \label{fig:phicontour}
\end{figure}

The aforementioned geometry, model, and parameters are used to run \emph{Virtual} fatigue crack growth experiments, in both air and hydrogen gas environments, and for various choices of loading frequency, load ratio and hydrogen gas pressure, mimicking the conditions of the laboratory tests. The model can simulate every load cycle and therefore naturally accommodate different load ratios and loading frequencies, as in the experiments. A representative example of the qualitative outcome of these computational experiments is given in Fig. \ref{fig:phicontour}, as computed for the case of a H$_2$ pre-charged sample exposed to a loading frequency of $f=1$ Hz, a hydrogen pressure of $p_\mathrm{H_2}=106$ MPa and a loading ratio of $R=0.1$. Fig. \ref{fig:phicontour}a depicts the crack growth process, through the contours of the phase field variable $\phi$. The crack front is defined to be in the middle of the diffuse interface ($\phi=0.5$). Fig. \ref{fig:phicontour}b shows hydrogen concentration contours, showing how hydrogen accumulates in areas of high hydrostatic stress, such as the moving crack tip. From the crack extension and the applied load, see Eq. (\ref{eq:DeltaK}), d$a$/d$N$ vs $\Delta K$ curves are obtained. 

\subsection{Predicting the role of hydrogen pressure, load ratio and loading frequency}

\begin{figure}[H]
\centering
\includegraphics[scale=0.13]{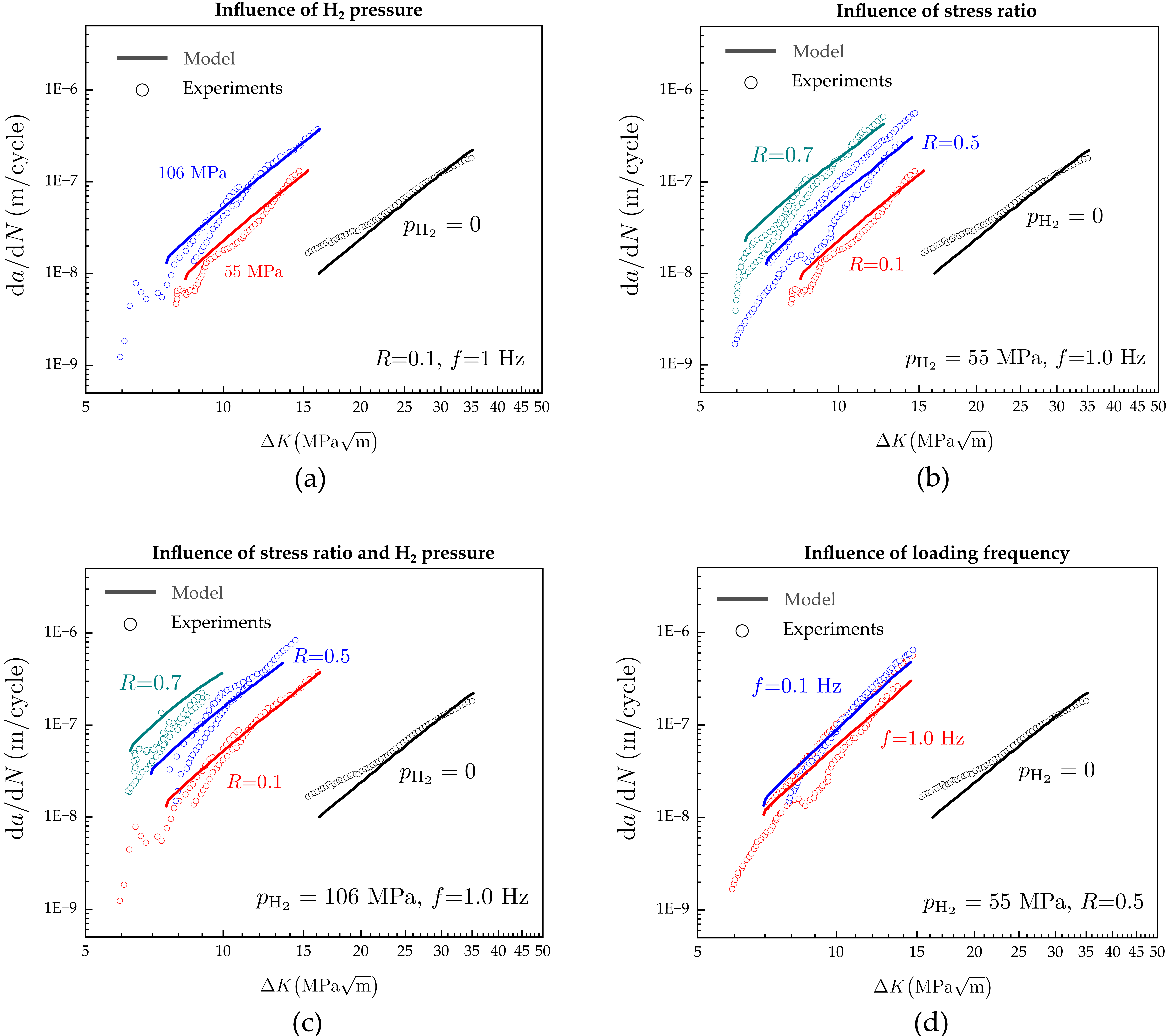}
\caption{Computational (lines) and laboratory (symbols) fatigue crack growth experiments for various hydrogen concentrations, load ratios and loading frequencies. The \emph{Virtual} experiments are shown to quantitatively capture the influence of (a) hydrogen pressure $p_\mathrm{H_2}$, (b) load ratio $R$, (c) load ratio $R$ and hydrogen pressure $p_\mathrm{H_2}$, and (d) loading frequency $f$. The experimental results are taken from Refs. \cite{San2017,bortot2023effect}.}
\label{fig:Comparison}
\end{figure}

The ability of the model to predict the behaviour of pressure vessel steels in the combined presence of hydrogen and cyclic loading is examined next. Recall that the model has been endowed with the material's fatigue characteristics (in air) and with the sensitivity of $G_c$ to hydrogen, yet no criteria or parameter has been defined to predict hydrogen-assisted fatigue. As shown in Fig. \ref{fig:Comparison}, the predictive abilities of the model are benchmarked against a wide range of $\Delta K$ values and scenarios, spanning different hydrogen pressures, load ratios and loading frequencies. The results exhibit a remarkable degree of agreement across all the conditions examined. 

Consider first, Fig. \ref{fig:Comparison}a, where the influence of the H$_2$ pressure is investigated. As stated above, changes in H$_2$ pressure result in different environmental hydrogen concentrations, as per Sievert's law (\ref{eq:Sievert}) and this is introduced into the simulation through the boundary conditions $C=C_{\text{env}}$. The results show that the model, calibrated to fit the response in air, is able to adequately \emph{predict} the experimental data for both $p_\mathrm{H_2}=55$ and $p_\mathrm{H_2}=106$ MPa. As in the experiments, these numerical calculations are obtained with a load ratio of $R=0.1$ and a loading frequency of $f=1$ Hz. Next, the load ratio is varied to assess the ability of the model to predict the interplay between stress ratio and hydrogen embrittlement susceptibility. The results, shown in Fig. \ref{fig:Comparison}b, reveal that the model quantitatively captures how fatigue crack growth rates become larger for higher $R$ values in the presence of hydrogen. The agreement is remarkable across the entire range, $R=0.1-0.7$. The numerical and experimental results shown in Fig. \ref{fig:Comparison}b were obtained for a loading frequency of $f=1$ Hz and a hydrogen pressure of $p_\mathrm{H_2}=55$ MPa, but experimental data for various $R$ values is also available for a pressure of $p_\mathrm{H_2}=106$ MPa. Thus, the ability of the model to predict fatigue crack growth rates across $p_\mathrm{H_2}$ and $R$ values can also be assessed - this is shown in Fig. \ref{fig:Comparison}c, where again a very good agreement with experiments is observed across the entire range of conditions considered. Finally, the role of the loading frequency is examined in Fig. \ref{fig:Comparison}d. Both the experiments and the numerical results show little sensitivity to the loading frequency. Overall, average trends in the experimental data suggest that fatigue crack growth rates are higher for lower loading frequencies, although the $f=0.1$ Hz and $f=1$ Hz curves overlap and cross each other for low $\Delta K$ values. The numerical results seem to reinforce this observation, with the $f=0.1$ Hz result being above the $f=1$ Hz one but with little differences being observed. While hydrogen-assisted fatigue crack growths are known to be sensitive to the loading frequency, the longer the loading cycle, the more time for hydrogen to accumulate in the process zone, a small sensitivity to the loading frequency is expected in the two extreme regimes (sufficiently fast or slow tests). Mapping the fatigue behaviour regimes that result from the interplay between material diffusivity and loading frequency is therefore key to optimising testing protocols and ensuring conservatism; this is the objective of the next subsection.

\subsection{The loading frequency conundrum}

\begin{figure}[H]
\centering
\includegraphics[scale=0.3]{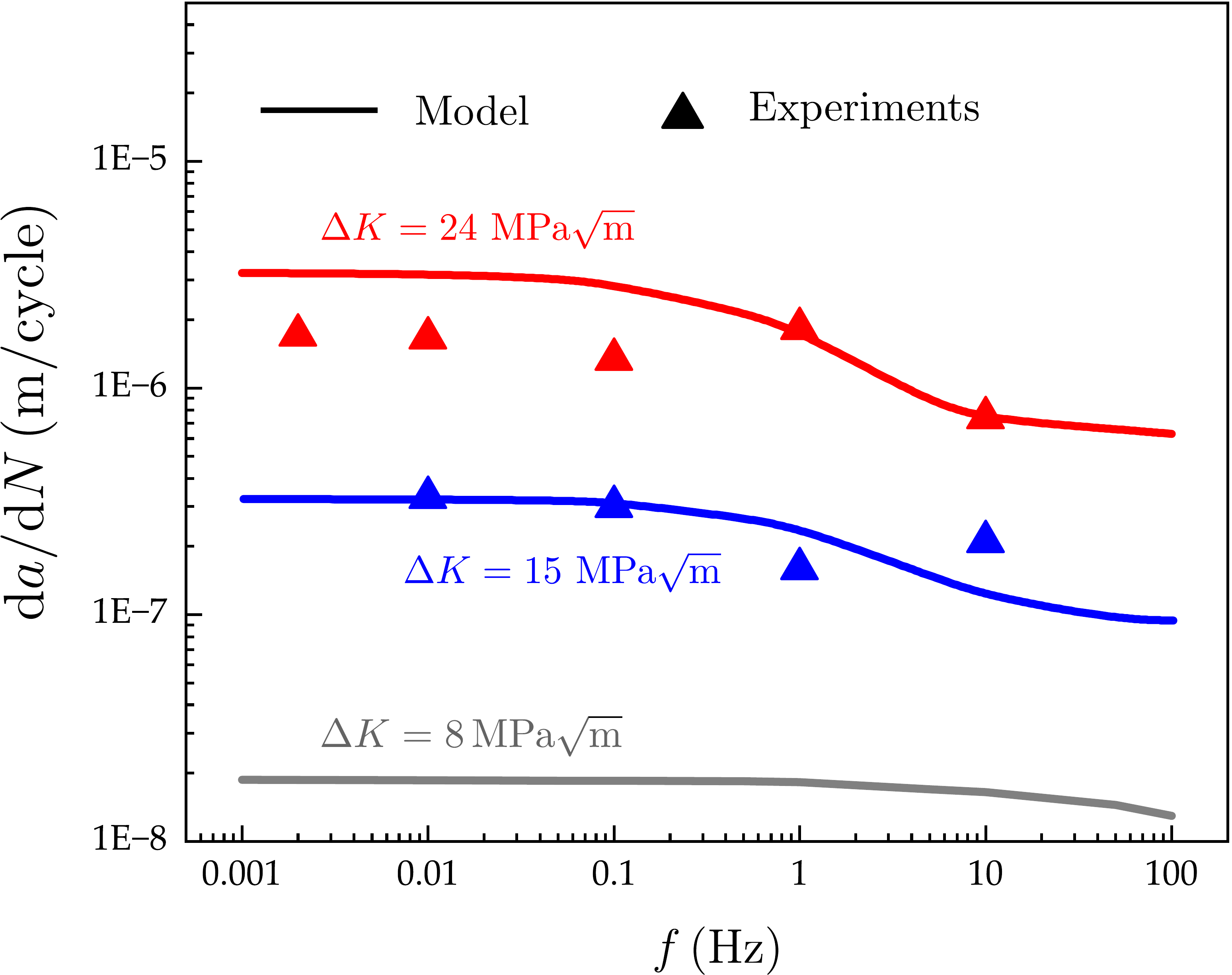}
\caption{Mapping the susceptibility of hydrogen-assisted fatigue crack growth rates to the loading frequency $f$. Computational predictions and experimental results are shown for three selected values of stress intensity factor range, $\Delta K$ values of 8, 15 and 24 $\mathrm{MPa\sqrt{m}}$. The load ratio is $R=0.1$ and the H$_2$ pressure equals $p_\mathrm{H_2}=103$ MPa. The experimental results are taken from Ref. \cite{somerday2015optimizing} and correspond to a similar pressure vessel steel (SA-372 Gr. J steel). The model quantitatively maps the three expected regimes of behaviour: sufficiently slow tests (left, high d$a$/d$N$ plateau), sufficiently fast test (right, low d$a$/d$N$ plateau), and intermediate regime of d$a$/d$N$ sensitivity to the loading frequency $f$.} \label{fig:dadNf}
\end{figure}

To unravel the interplay between hydrogen transport, loading frequency and fatigue crack growth rates in pressure vessel steels, numerical experiments are conducted over a wide range of loading frequencies, from $f=0.001$ Hz to $f=100$ Hz. There is a lack of experimental data at low frequencies due to the challenges associated with conducting tests at those conditions (large time scales, leaks). However, experimental data for a wide range of loading frequencies is available for SA-372 Gr. J steel, a similar material with $\sigma_y=760$ MPa (see Ref. \cite{somerday2015optimizing}), and this is used to further benchmark model predictions. The results obtained are shown in Fig. \ref{fig:dadNf}, for three selected values of the stress intensity factor range: $\Delta K =8$ MPa$\sqrt{\text{m}}$, $\Delta K =15$ MPa$\sqrt{\text{m}}$ and $\Delta K =24$ MPa$\sqrt{\text{m}}$. The load ratio is $R=0.1$ and the H$_2$ pressure equals $p_\mathrm{H_2}=103$ MPa. Three observations should be particularly emphasised. Firstly, the model is shown to accurately capture the three expected regimes of behaviour: (i) a ``fast test'' regime where hydrogen does not have time to diffuse and d$a$/d$N$ values are lowest, (ii) a ``slow test'' regime, where hydrogen has sufficient time to diffuse to the fracture process zone region and d$a$/d$N$ values are highest, and (iii) an intermediate regime, where fatigue crack growth rates are susceptible to the loading frequency. Among others, the range of loading frequencies that this intermediate regime spans is dependent on the $\Delta K$ value and, mainly, the diffusivity of the material. Secondly, computational and experimental results appear to be in relatively good agreement, despite the significant scatter observed in the latter. Some differences are observed in the ``slow test'' regime (very low $f$) for the highest load $\Delta K$. These could be potentially related to the challenges intrinsic to low loading frequency tests (e.g., a higher propensity for leaks). Thirdly, model results show that the impact of loading frequency becomes more significant as $\Delta K$ increases. The sensitivity to $\Delta K$ is notable and can be explained by the dependency of the peak hydrostatic stress (and thus peak $C$) on $K$ and by the interplay between hydrogen transport and crack growth rate, with the latter increasing significantly with $\Delta K$. All virtual experiments conducted in the mid-low $\Delta K$ regime show a lower sensitivity to the loading frequency and a wider range of conservative test frequencies. A loading frequency of 1-10 Hz appears to be sufficient if $\Delta K$ is on the order of 8 MPa$\sqrt{\text{m}}$. On the other hand, for values of the stress intensity factor range as high as $\Delta K =24$ MPa$\sqrt{\text{m}}$, our results show that one has to reduce the loading frequency to roughly 0.1 Hz to be sufficiently close to the ``slow test'' regime and therefore conservative. Tests conducted at $f=1$ Hz (a common habit), appear to under-predict by a factor of 2 the fatigue crack growth representative of engineering applications (where $f \ll 1$ Hz). Nevertheless, the high $\Delta K$ regime is arguably of less technological significance as most pressure vessels for H$_2$ storage are currently designed to operate at mid to low $\Delta K$ values.

\subsection{The role of pre-loading hydrogen exposure time}

\begin{figure}[H]
\centering
\includegraphics[scale=0.2]{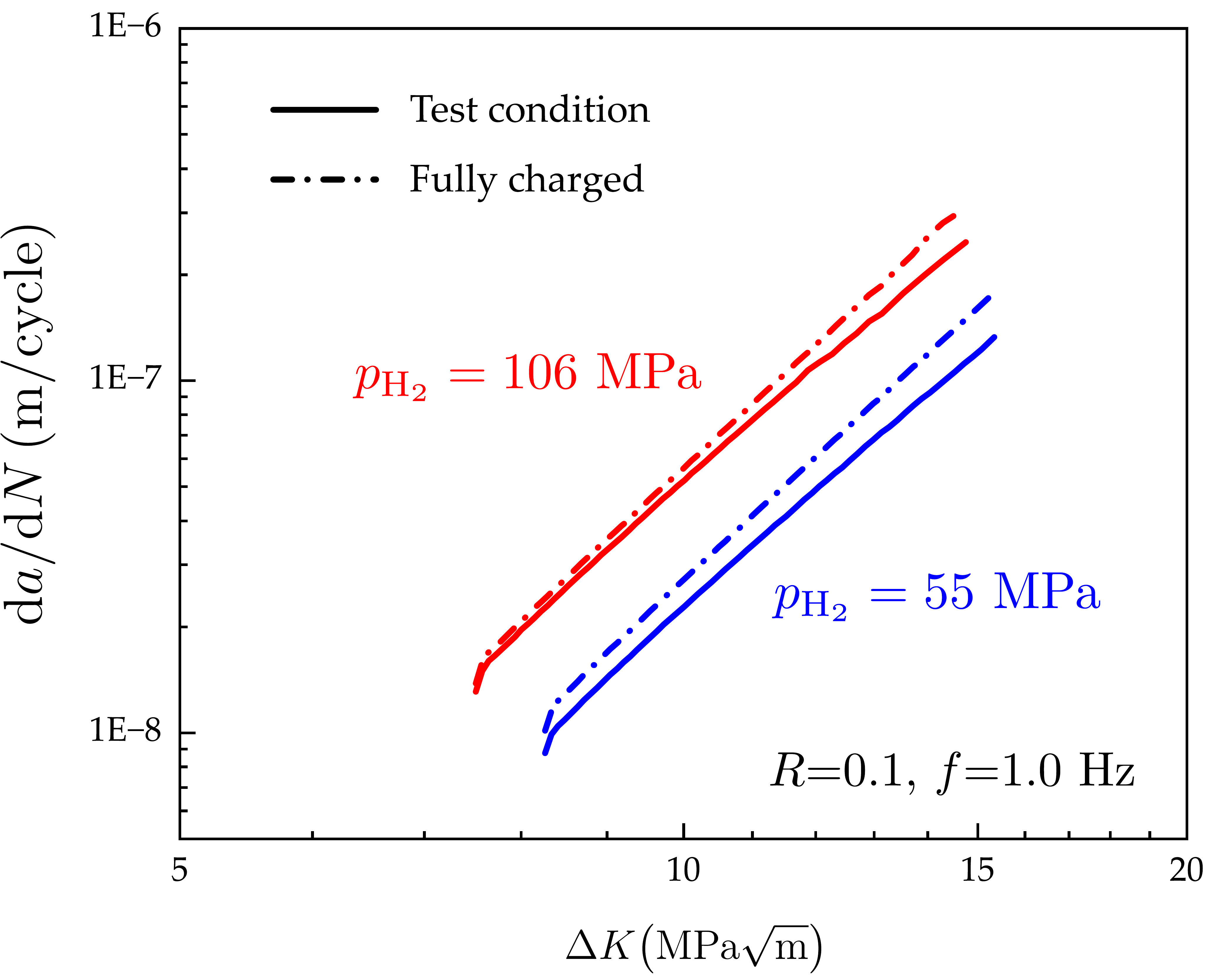}
\caption{Gaining insight into the influence of H$_2$ pre-charging before mechanical loading. Computational predictions of fatigue crack growth curves $\text{d}a/\text{d}N$ as a function of the stress intensity factor range $\Delta K$ for typical tests conditions (24 h of H$_2$ exposure time before mechanical loading) and the scenario where the sample is completely saturated with hydrogen before mechanical testing. Results are obtained for a loading frequency of $f=1$ Hz, a load ratio of $R=0.1$ and two hydrogen pressures, $p_\mathrm{H_2}=55$ and $p_\mathrm{H_2}=106$ MPa.} \label{fig:Fullcharge}
\end{figure}

Finally, the model is used to gain insight into the influence of the so-called `soak time', the period when the sample is exposed to hydrogen before mechanical loading starts. In the experiments considered so far, this time was approximately 24 hours, as this is the time that it takes to stabilise the load cell. With state-of-the-art Linear Variable Differential Transformer (LVDT) load cells this can be reduced to minutes. In general, samples are not typically exposed to a hydrogen-containing environment prior to charging to reduce the testing time and to emulate the scenario where the material is protected from hydrogen uptake until a fatigue crack develops (e.g., because of the presence of an oxide or an engineering coating). However, in components experiencing long-term exposure to hydrogen, one can readily envisage a situation where the material has absorbed hydrogen before the onset of fatigue crack growth. Consequently, we use our model to quantify the deviation between the fast, no pre-charging condition and the potentially more technologically relevant scenario of hydrogen saturation before mechanical testing. To this end, calculations are conducted with the following initial conditions: $C(t=0)=C_\mathrm{env} \, \forall \, \textbf{x}$. That is, a uniform hydrogen concentration (the one corresponding to the relevant applied pressure) throughout the sample. The results obtained are shown in Fig. \ref{fig:Fullcharge}. The numerical experiments were conducted for a load ratio of $R=0.1$, a loading frequency of $f=1$ Hz and two H$_2$ pressures: $p_\mathrm{H_2}=55$ MPa and $p_\mathrm{H_2}=106$ MPa. The obtained fatigue crack growth curves show a very small effect at low $\Delta K$ values but a larger deviation at high $\Delta K$ values. Also, differences appear to be slightly smaller with increasing $p_\mathrm{H_2}$, as the toughness sensitivity to changes in hydrogen content is smaller for high $C$ values (see Fig. \ref{fig:KthvsH}). The qualitative trends observed are not surprising, as in the high $\Delta K$ regime the crack is growing into a region that was free of hydrogen in the conventional test condition. This non-uniform distribution of the material toughness is much less pronounced in the fully charged numerical experiments, leading to higher d$a$/d$N$ values. Less sensitivity to the loading frequency is also expected in the pre-charged condition, although there will still be a dependency due to the interplay between hydrostatic stresses and hydrogen transport. 
 
\subsection{Towards a model incorporating a hydrogen-fatigue interplay}
\label{app:AppendixX}

\begin{figure}[H]
\centering
\includegraphics[scale=0.28]{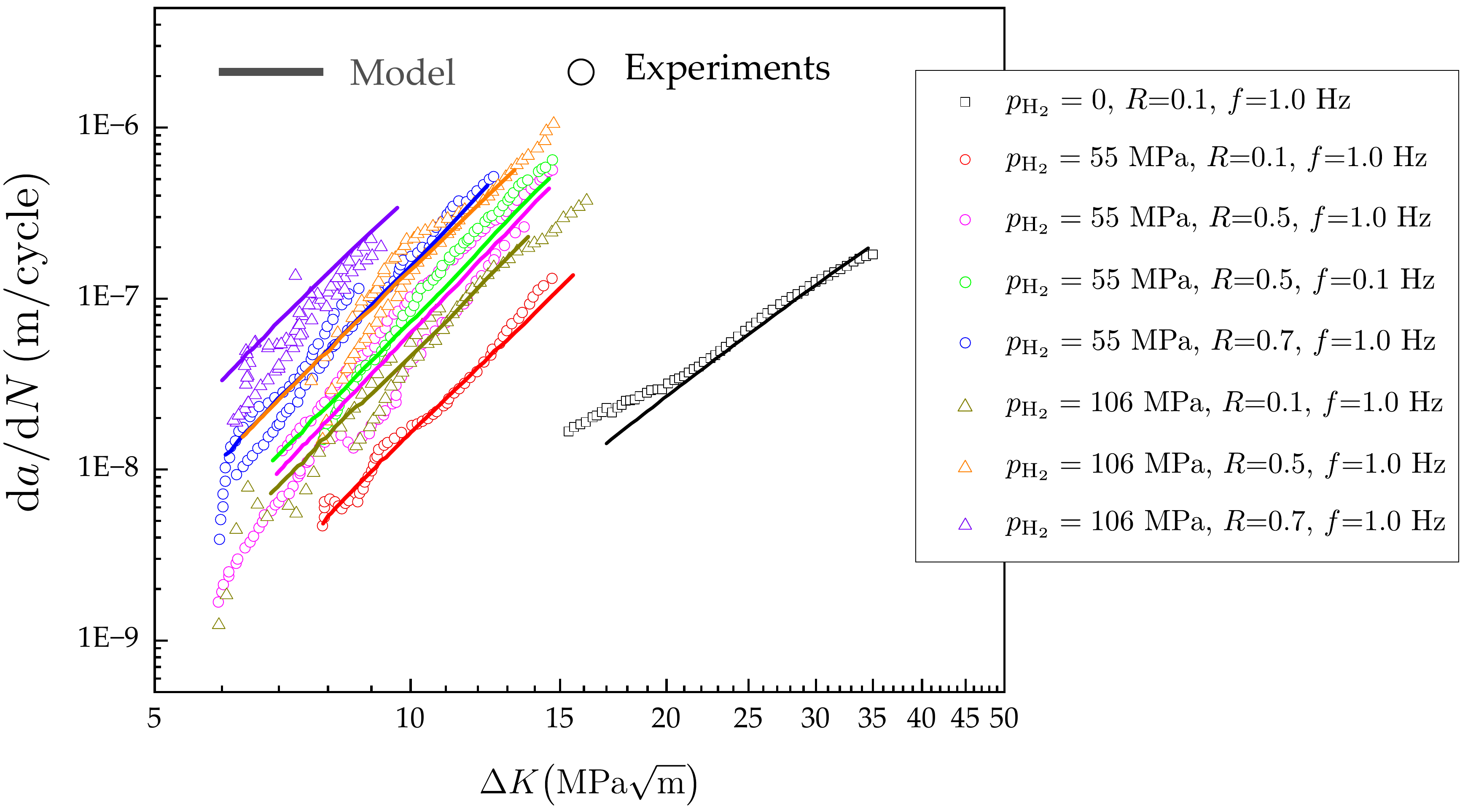}
\caption{Enhancing predictions by accounting for the hydrogen-cyclic damage interplay. Computational (lines) and laboratory (symbols) fatigue crack growth experiments for various hydrogen concentrations, load rations and loading frequencies. Different to the rest of the manuscript, the fatigue parameter $n$ is chosen to take a different value for hydrogen environments ($n=1.9$). Predictions show a very good agreement with experiments for all the values of stress intensity factor range $\Delta K$, loading frequency $f$, load ratio $R$ and hydrogen pressure $p_\mathrm{H_2}$.} \label{fig:Expdata_n}
\end{figure}

The model proposed is capable of accurately predicting fatigue crack growth rates based solely on one element of coupling between the damage process and hydrogen: the degradation of fracture toughness with increasing hydrogen content. This assumes that hydrogen only reduces the fracture resistance of the material and has no effect on the ability of a material to develop damage due to cyclic loading. This is arguably a sensible assumption, as the impact of hydrogen in fatigue is mostly through an enhancement of fatigue crack growth rates, and the results obtained reveal that this assumption enables establishing a model capable of delivering accurate predictions with minimal calibration. However, changes in S-N curve behaviour with increasing hydrogen content have been reported \cite{Matsunaga2015}, which suggests that hydrogen can also accelerate the nucleation of fatigue cracks. While the present model can naturally capture this through the relationship between strength and (hydrogen-degraded) toughness, the extent to which this agrees quantitatively with experiments has not been thoroughly investigated yet. Thus, an argument could be made for the development of a more flexible model, capable of accounting for a hydrogen-fatigue interplay through the addition of an additional parameter. To showcase the potential improvement in accuracy that such a model would bring, fatigue crack growth curves are re-calculated considering a different $n$ value for hydrogen-containing environments. Specifically, Fig. \ref{fig:Expdata_n} shows the results obtained using $n=1.9$ for $p_\mathrm{H_2}>0$ (with $n=1$ for $p_\mathrm{H_2}=0$). A slightly better agreement is attained, relative to the results presented in Fig. \ref{fig:Comparison}.

\section{Concluding remarks}
\label{Sec:Conclusions}

We have presented a phase field-based computational framework for predicting hydrogen-assisted fatigue in metallic materials. The model combines: (i) a variational phase field description of the solid-interface, (ii) an extended version of Fick’s law to describe the hydrogen transport within the metallic lattice, (iii) a generalised definition of fatigue damage accumulation that can accommodate the role of the stress ratio, the fatigue threshold and arbitrary S-N and Paris law slopes, and (iv) the degradation of fracture toughness due to both hydrogen embrittlement and cyclic loading. This novel framework is employed to conduct \emph{virtual} fatigue crack growth experiments for various loading frequencies, load ratios and hydrogen gas pressures. The model delivers quantitative hydrogen-assisted fatigue predictions without any parameter fitting or `ad hoc' criterion, taking as input solely the fatigue characteristics of the material in the absence of hydrogen and the dependency of the material toughness on hydrogen concentration. Several findings can be emphasised:\\

\begin{itemize}
    \item The model can capture the influence that hydrogen pressure, stress ratio, and loading frequency have on hydrogen-assisted fatigue crack growth. Moreover, the results obtained are in excellent agreement with experimental data, revealing remarkable predictive abilities and suggesting that the fatigue crack growth behaviour in hydrogen-containing environments can be rationalised solely by resolving the interplay between loading cycle duration, stress-driven hydrogen transport and fracture toughness sensitivity to hydrogen.\\

    \item Model predictions are used to map regimes of loading frequency behaviour in pressure vessel steels. Two limit regimes are identified whereby fatigue crack growth rates are insensitive to the loading frequency; sufficiently `fast tests', where hydrogen does not have time to diffuse and d$a$/d$N$ values are the lowest, and sufficiently `slow tests', where hydrogen has time to achieve a distribution close to the steady state and d$a$/d$N$ values are the highest. The use of a loading frequency of $f=1$ Hz, as typically done in experiments, would under-predict fatigue crack growth rates by a factor of 1.5-2 in the mid-high $\Delta K$ region. However, a loading frequency of $f=1$ Hz is appropriate for $\Delta K$ values smaller than 15 MPa$\sqrt{\text{m}}$, a regime that is currently of more technological importance. \\

    \item Insight is gained into the implications of pre-charging the samples prior to testing. Current testing protocols typically have a 24 h of H$_2$ exposure before the onset of cyclic loading; this was found to have a moderate effect relative to the more conservative scenario of exposure until sample saturation. Differences increase with $\Delta K$, with fatigue crack growth rates being roughly 2 times higher for stress intensity factor ranges of $\Delta K=15$ MPa$\sqrt{\text{m}}$.\\  
\end{itemize}

The framework presented can provide quantitative predictions of hydrogen-assisted fatigue behaviour from static fracture data at various hydrogen pressures and one fatigue crack growth curve in air. This not only enables conducting fast and cheap numerical experiments, within and beyond the conditions achievable by laboratory experiments, but also enables performing \emph{Virtual Testing} of engineering components as the model can handle arbitrary geometries (2D, 3D) over large spatial and temporal scales. 

\section*{CRediT authorship contribution statement}
\noindent \textbf{C. Cui:} Investigation, Methodology, Programming, Writing – original draft, Writing – review \& editing. \textbf{P. Bortot:} Investigation, Methodology, Funding acquisition, Writing – review \& editing. \textbf{M. Ortolani:} Investigation, Writing – review \& editing. \textbf{E. Mart\'{\i}nez-Pa\~neda:} Investigation, Conceptualization, Methodology, Funding acquisition, Supervision, Writing – review \& editing. 

\section*{Declaration of Competing Interest}
The authors declare that they have no known competing financial interests or personal relationships that could have appeared to influence the work reported in this paper.

\section*{Acknowledgements}
\label{Sec:Acknowledgeoffunding}
 
The authors acknowledge financial support from TENARIS through the R\&D project ``A Virtual Platform to Assess the Integrity of Hydrogen-Containing Components''. C. Cui acknowledges financial support from the UKRI Horizon Europe Guarantee MSCA Postdoctoral Fellowship (grant EP/Y028236/1). Emilio Mart{\' i}nez-Pa{\~ n}eda additionally acknowledges financial support from UKRI’s Future Leaders Fellowship programme [grant MR/V024124/1]. The authors also acknowledge computational resources and support provided by the Imperial College Research Computing Service ( \url{https://doi.org/10.14469/hpc/2232}).

\section*{Supplementary material}
\label{app:Supplementary material}

Supplementary material associated with this article can be found in the online version.


\bibliographystyle{elsarticle-num}
\biboptions{sort&compress}

\end{document}